\documentclass[twocolumn,showpacs,preprintnumbers,prb,amsmath,amssymb]{revtex4}

\usepackage{graphicx}%
\usepackage{dcolumn}
\usepackage{amsmath}
\usepackage{color}

\makeatletter
\def\btt#1{\texttt{\@backslashchar#1}}%
\DeclareRobustCommand\bblash{\btt{\@backslashchar}}%
\makeatother

\begin{document}

\title{Anomalous properties in the low-carrier ordered phase of PrRu$_4$P$_{12}$: Consequence of hybridization between conduction
    and Pr 4$f$ electrons}

\author{S.R.~Saha}

\altaffiliation{Present Address: Center for Nano Physics and Advance
Material, Department of Physics, University of Maryland, MD 20742,
USA. Email: srsaha@umd.edu}

\author{H.~Sugawara}
\altaffiliation{Present Address: Faculty of Integrated Arts and
Sciences, The University of Tokushima, Tokushima 770-8502, Japan.}

\author{T. Namiki}
\author{Y.~Aoki}
\author{H.~Sato}

\address{Department of Physics, Tokyo Metropolitan University,
Hachioji, Tokyo 192-0397, Japan}

\date{\today}

\begin{abstract}
~~~~~The low-carrier ordered phase below the metal-non-metal
transition temperature $T_{\rm MI}\simeq63$~K of PrRu$_4$P$_{12}$ is
explored by probing magnetoresistance, magnetic susceptibility,
thermoelectric power, and Hall effect on high quality single
crystals. All the measured properties exhibit the signature of
decimation of the Fermi surface below $T_{\rm MI}$ and anomalous
behaviors below 30 K including a large thermoelectric power $\sim
-200\mu$V/K and a giant negative magnetoresistance (93\% at $\sim$
0.4 K). The results indicate an additional structure below 30 K and
a semimetal-like ground state. The observed anomalous behaviors are
most likely associated with the novel role of $c$-$f$ hybridization
between conduction electrons and Pr 4$f$ electrons, whose
crystalline electric field level schemes show drastic change below
$T_{\rm MI}$.
\end{abstract}


\maketitle


\section{INTRODUCTION}
The filled-skutterudite compounds $R$Tr$_4$Pn$_{12}$ ($R$= rare
earth, Tr= Fe, Ru, Os; and Pn=pnictogen) have attracted much
attention for exhibiting unusual physical properties and their
prospect in thermoelectric
applications.~\cite{sales,sekine1,torikak,sato1,bauer,suga1,aoki1,aoki2,suga2,gumeniuk}
Particularly interesting are some Pr-based
skutterudites,~\cite{aokireview} which are unlike the Pr metal and
ordinary Pr compounds that possess well-localized 4$f$ electronic
states. The exotic behaviors observed in these compounds are
believed to be associated with strong electron correlation or
$c$-$f$ hybridization between conduction and Pr 4$f$
electrons.~\cite{bauer,sato1,suga1,aoki1,aoki2,suga2,aokireview} In
this context, PrRu$_4$P$_{12}$ deserves further attention, since the
mechanism of its exotic metal-insulator (M-I)
transition~\cite{sekine1} at $T_{\rm MI}\simeq60$~K and the ordered
phase remain as mysterious puzzles, despite several studies.
\cite{sekine1,sekine2,lee1,curnoe,harima1,saha1,hao1}

Multipole order cooperating with the Fermi-surface (FS) nesting
resulting in the charge density wave (CDW) has been suggested by a
band-structure calculation as one possible mechanism of the M-I
transition.\cite{harima3} A structural transition below $T_{\rm MI}$
with the doubling of the unit cell and a change in space group from
$Im\overline 3$ above $T_{\rm MI}$ to $Pm\overline 3$ below $T_{\rm
MI}$ has been observed.\cite{lee1} We have succeeded to measure the
de Haas-van Alphen (dHvA) effect clarifying the FS topology in
LaRu$_4$P$_{12}$ and found that its FS is similar to the theoretical
FS of PrRu$_4$P$_{12}$~(Ref.~\onlinecite{saha3}) with the nesting
instability. However, LaRu$_4$P$_{12}$ does not show any M-I
transition, suggesting that the $c$-$f$ hybridization between Pr
4$f$ and conduction electrons is essential for the M-I transition in
PrRu$_4$P$_{12}$.\cite{saha3} The inelastic neutron scattering (INS)
experiment indicates the presence of strong $c$-$f$ hybridization
above $T_{\rm MI}$, while below $T_{\rm MI}$ the strength of
hybridization reduces and Pr 4$f$ electrons attain almost localized
nature.\cite{iwasa} The INS results also suggest that the
crystalline electric field (CEF) ground state is nonmagnetic
$\Gamma_1$ singlet across $T_{\rm MI}$ both at the body center
(Pr$_1$) and at the cubic corner (Pr$_2$) of the unit cell, while
Pr$_2$ sites show an abrupt change to magnetic $\Gamma_4^{(2)}$
triplet below around 30 K keeping Pr$_1$
unchanged.\cite{iwasa,iwasa1} Recently, the M-I transition has been
ascribed theoretically to an antiferro-hexadecapole order or scalar
order without breaking the local symmetry at the
Pr-site.\cite{takimoto,Kiss2008} However, a recent photoemission
study claims that there is no notable change in $c$-$f$
hybridization strength across the M-I transition.\cite{yamasaki}

We have succeeded in growing high-quality single crystals of
PrRu$_4$P$_{12}$ and its reference compound LaRu$_4$P$_{12}$, which
has no 4$f$-electron. In this paper, we report on the study of their
transport and magnetic properties which documents the decimation of
the Fermi surface below $T_{\rm MI}$ and anomalous behaviors at low
temperatures in PrRu$_4$P$_{12}$. Indications of
temperature-dependent carrier scattering from Pr-sites and a direct
coupling between 4$f$ and conduction electrons have been observed in
PrRu$_4$P$_{12}$.

\section{EXPERIMENT}
Single crystals of PrRu$_4$P$_{12}$ and LaRu$_4$P$_{12}$ were grown
by the tin-flux method.\cite{torikak} The raw materials were 4N
(99.99 \% pure)-La, -Pr, -Ru, 6N-P, and 5N-Sn. The single
crystalline nature has been checked by the back Laue x-ray
scattering technique. X-ray powder-diffraction (XRD) experiment
confirmed the filled-skutterudite structure and absence of any
impurity phases. The results are also consistent with the
calculation and the previous XRD report on the high-pressure grown
polycrystal.\cite{sekine1} We have succeeded in observing the de
Haas-van Alphen effect in LaRu$_4$P$_{12}$,\cite{saha3} which
indicates the high sample quality and also indirectly assures the
high quality of PrRu$_4$P$_{12}$ single crystals grown by the same
manner. Electrical resistivity, magentoresistance, and  Hall effect
were measured by the standard dc four probe method. All these
measurements were performed in a top loading $^3$He cryostat, which
can be cooled down to $\sim$0.3 K, equipped with a superconducting
magnet capable of generating magnetic fields up to 16 T.  The
temperature dependence of the Hall effect was measured in a $^3$He
cryostat down to $\sim $0.4 K under a magnetic field of 1.5 T
generated by a dc current sweep magnet. The thermoelectric power
data have been taken by the differential method using a Au-Fe
(0.07\%)-chromel thermocouple in a $^4$He cryostat cooled down to
$\sim 1.5$~K. The magnetic susceptibility was measured by a Quantum
Design superconducting quantum interference device (SQUID)
magnetometer.

\section{RESULTS}
\subsection{Electrical resistivity and magnetoresistance}
\begin{figure}[hbtp]
\begin{center}\leavevmode
\includegraphics[width=8.6cm]{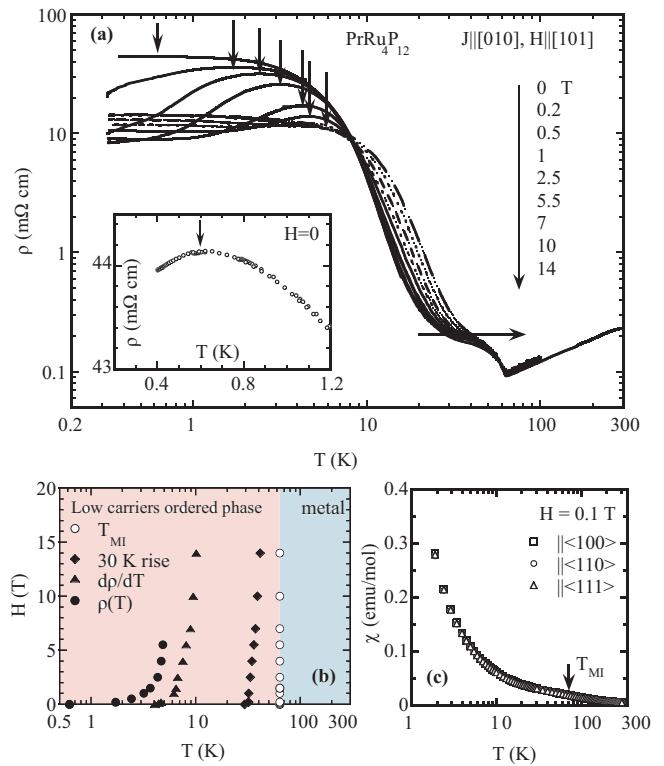}
\caption{(a). Temperature dependence of electrical resistivity,
$\rho (T)$, in PrRu$_4$P$_{12}$ under several magnetic fields. The
short arrows indicate the shift of the position of the broad peak in
$\rho (T)$ with increasing applied fields. The horizontal and the
long vertical arrows indicate the shift in the $\rho(T)$ curves with
increasing fields. The inset shows an expanded view of the faint
peak around 0.6 K, as indicated by the short arrow, in $\rho (T)$
presented in the linear scales for $H=0$. (b) The phase diagram
determined by the temperatures where the $\rho(T)$ shows peaks
(filled circles), $d\rho(T)/dT$ shows negative peaks (filled
triangles), $\rho$ shows upturn below the shoulder around 30 K
(filled diamonds), and the kink at $T_{\rm MI}$ in $\rho$($T$) (open
circles) for different values of $H$. The shaded areas with two
different colors/intensities distinguish the normal metallic state
and the low-carrier ordered phase below $T_{\rm MI}$. (c)
Temperature dependence of magnetic susceptibility $\chi (T)$ in
single crystalline PrRu$_4$P$_{12}$ for the fields along three
different crystallographic directions.} \label{rho_T}
\end{center}
\end{figure}

\begin{figure}[hbtp]
\begin{center}\leavevmode
\includegraphics[width=8.6cm]{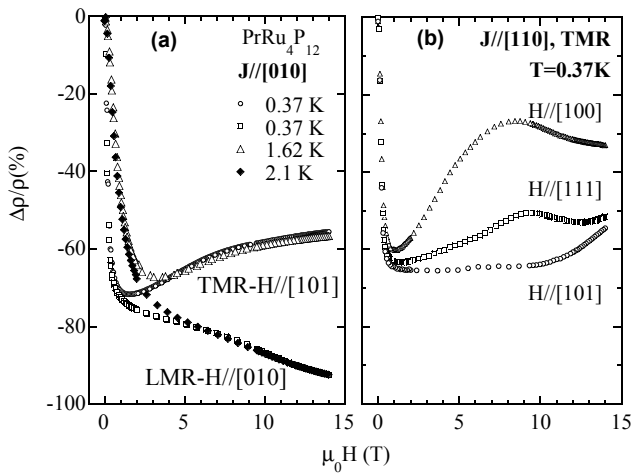}
\caption{(a). Field dependence of isothermal magnetoresistance in
PrRu$_4$P$_{12}$ for both transverse (TMR) and longitudinal (LMR)
geometries with the current $J\parallel [010]$. TMR for $J\parallel
[110]$ are given in panel (b).} \label{MR-H}
\end{center}
\end{figure}

Figure~\ref{rho_T}(a) shows the temperature ($T$) dependence of
resistivity ($\rho$) in PrRu$_4$P$_{12}$ under the magnetic field
($H$) from 0 to 14 T. For $H=0$, $\rho$ decreases with $T$ like
normal metals down to 63 K below which it increases with decreasing
$T$ showing a kink around 63 K that traces the so called M-I
transition. After showing a shoulder around 30 K $\rho$ rapidly
increases below $\sim$ 10 K and then tends to saturate resulting a
faint peak around 0.6 K as pointed by the arrow in the inset.
Overall $\rho(T)$ is similar to that of polycrystal grown under high
pressure (Ref.~\onlinecite{sekine1}). However, $\rho$(at 1.5
K)/$\rho$(at 300 K) is one order higher than that in the
polycrystal,\cite{sekine1} indicating the high quality of the
present single crystals. No divergence in $\rho (T)$ and the
decrease of resistivity with $T$ below 0.6 K indicate the ground
state of PrRu$_4$P$_{12}$ as a low carrier semimetal rather than an
insulator. There is no considerable effect of $H$ up to 14 T on
$\rho$ across $T_{\rm MI}$ and on the values of $T_{\rm MI}$, which
is consistent with the high field specific heat
measurement.\cite{sekine2} In contrast, the applied external
magnetic field ($H$) causes a change in $\rho$ below $\sim$30 K
followed by a suppression of $\rho$ below $\sim$10 K. The faint peak
around 0.6 K at $H=0$ shifts to higher $T$ with increasing $H$ as
indicated by the positions of the short arrows and then smears out
above $\sim$6 T. Figure~\ref{rho_T}(b) shows the $H vs. T$ phase
diagram that gives the $H$ dependence of the peaks of $\rho(T)$ and
$d\rho (T)/dT$. The peak in $\rho(T)$, indicating this is not a
simple insulator, may be caused by the strong correlation effect
with the triplet CEF ground state of Pr. The triplet splits in
applied fields due to Zeeman effect and the energy splitting of the
triplet increases with fields. Therefore, the observed shift in the
peak in $\rho(T)$ to higher temperatures in $H$ is probably
associated with the Zeeman splitting of the triplet. The drastic
effect of $H$ suggests that the magnetic state is changing below
$\sim$30 K in PrRu$_4$P$_{12}$. In contrast to a kink in $\rho(T)$,
there is no distinct anomaly around $T_{\rm MI}$ in the temperature
dependence of magnetic susceptibility, $\chi (T)$, as shown in
Fig.~\ref{rho_T}(c), which is consistent with the results on the
polycrystal,\cite{sekine1} suggesting a non magnetic origin of
$T_{\rm MI}$. The anisotropy in $\chi(T)$ measured under 0.1 T field
applied along different crystallographic directions is small.
%
%
Figures~\ref{MR-H}(a) and \ref{MR-H}(b) show the field dependence of
several isothermal magnetoresistance $\Delta
\rho=[\rho(H)-\rho(H=0)]/\rho(H=0)$ in both transverse and
longitudinal geometries in PrRu$_4$P$_{12}$. Surprisingly, $\rho$
sharply drops resulting in a negative giant ($\sim70\%$)
magnetoresistance (GMR) under $H\leq$0.5 T in both geometries among
which the longitudinal magnetoresistance (LMR) attains a value as
high as $\sim93\%$ in 14 T at 0.37 K. In contrast, transverse
magnetoresistance (TMR) in LaRu$_4$P$_{12}$ is positive\cite{saha3}
and increases almost linearly with $H$, indicating that the negative
GMR in PrRu$_4$P$_{12}$ is associated with the 4$f$-electron state.
Magnetic field dependence of magnetization $M(H)$ at 60 mK, measured
on the same single crystal on which $\chi (T)$ data were taken,
shows a sharp increase at $\sim$ 0.5 T, although no magnetic order
was detected down to 60 mK.~\cite{sakakibara} The sharp increase in
$M(H)$ was ascribed to the contribution from the CEF triplet ground
state of the 50\% Pr ions. Our recent specific heat measurement
confirms no magnetic order down to $\simeq$ 150 mK; the molar
entropy of $(Rln3)/2$ at 10 K is consistent with the triplet CEF
ground state for one of the two Pr-sites.\cite{aoki3} Despite the
little anisotropy in $M(H)$ at high fields\cite{sakakibara} and in
$\chi (T)$, magnetoresistance is, interestingly, anisotropic at
higher fields. There may be two components in the magnetoresistance;
one is isotropic and dominating in low fields, while the other is
anisotropic electronic response, possibly related with CEF, which is
playing a role at higher $H$.
%

\subsection{Thermoelectric power}
\begin{figure}[hbtp]
\begin{center}\leavevmode
\includegraphics[width=8.6cm]{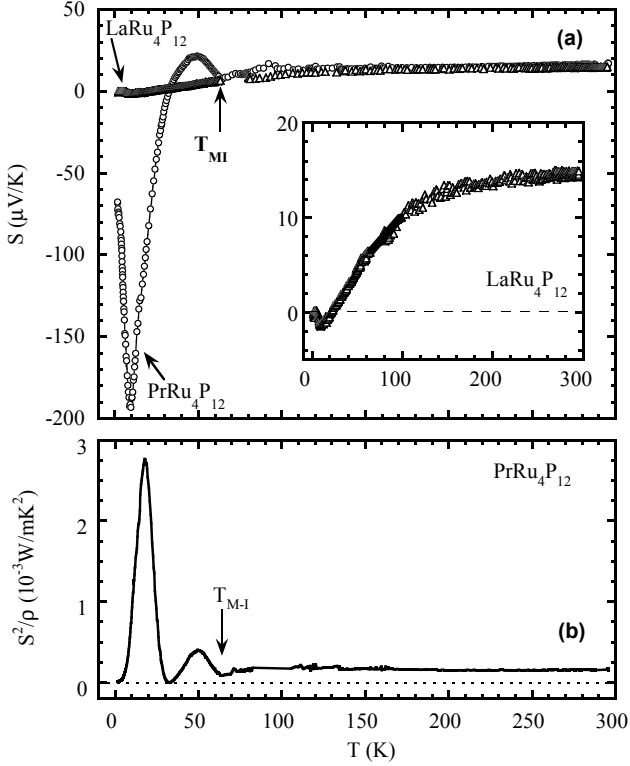}
\caption{. Temperature dependence of thermoelectric power $S(T)$ in
PrRu$_4$P$_{12}$ and LaRu$_4$P$_{12}$. Data were taken on arrays of
more than one single crystal due to size constraints. The inset
shows $S(T)$ in LaRu$_4$P$_{12}$ in an expanded $y$-axis view. (b)
The temperature dependence of thermoelectric power factor $S^2/\rho$
in PrRu$_4$P$_{12}$.} \label{Thermo_T}\end{center}\end{figure}

Figure~\ref{Thermo_T}(a) shows a temperature dependence of
thermoelectric power $S(T)$ in PrRu$_4$P$_{12}$ and
LaRu$_4$P$_{12}$. At the room temperature (RT) the equal value of
$S\sim 15 \mu$ V/K in both compounds indicates similar electronic
states at the vicinity of the Fermi level $E_{\rm F}$. The large
absolute value of $S$ compared to that of the simple metals may be
ascribable to the large contribution from the 4$d$-bands of Ru
and/or 5$p$-bands of P near the Fermi level.\cite{saha3} In
LaRu$_4$P$_{12}$, $S$ is positive at RT and it gradually decreases
with temperature (see the inset). There is no anomaly in $S(T)$
around 60 K, while $S$ changes sign below $\sim$ 20 K. This type of
temperature dependence might be due to a combination of diffusion
and phonon drag contribution. After showing a minimum around $\sim$
10 K, $S$ goes sharply to zero below 7 K due to
superconductivity.\cite{saha3} In PrRu$_4$P$_{12}$, $S$ shows a
similar temperature dependence down to $\sim 65$ K below which it
increases across $T_{\rm MI}$ tracing the M-I transition, indicating
a change in the Fermi surface. After showing a maximum at $\sim 50$
K $S$ changes sign below $\sim 30$ K and then exhibits a minimum at
$\sim$ 10 K with a dramatically large negative value of $\sim -200
\mu$ V/K. Such a large negative value of $S$ was also reported in
PrFe$_4$P$_{12}$ in its ordered state below 6.5 K\cite{Pourret}
where the low carrier density was confirmed. In many correlated
metals, the absolute value of the dimensionless ratio
$q=SN_{A}e/T\gamma$ (where $\gamma$ is the linear coefficient of the
electronic specific heat, $e$ is the elementary charge, and $N_{A}$
is the Avogadro number) is of the order of unity.\cite{Behnia} The
slope of thermopower below 1.8 K is large ($S/T \simeq -43 \mu$
V/K$^2$ at 1.6 K) in PrRu$_4$P$_{12}$. Following a similar
assumption to Ref.~\onlinecite{Pourret} and assuming a value of
$\gamma \sim$0.1 J/K$^2$ mol in PrRu$_4$P$_{12}$, since a precise
value of $\gamma$ at low $T$ is not yet determined,\cite{aoki3} one
can get a value of $q \simeq -43$ at 1.6 K in PrRu$_4$P$_{12}$. This
value is of similar order to $q \simeq -53$ at $\sim$0.1 K in
PrFe$_4$P$_{12}$,\cite{Pourret} suggesting a presence of large
correlation in the non-metallic state of PrRu$_4$P$_{12}$. In order
to interpret the $S(T)$ behavior in PrRu$_4$P$_{12}$, one should
consider the diffusion thermopower represented by an energy
derivative of conductivity $\sigma(\varepsilon)$ as
$S=-(\pi^2k_B^2T/3e\sigma)[d\sigma(\varepsilon)/d\varepsilon]_{\varepsilon_F}$,\cite{dugdale}
where $d\sigma(\varepsilon)/d\varepsilon$ depends on carrier numbers
and scattering probability. In the high-mobility semiconductor a
large value of $S$ could be expected at low $T$ where the carrier
numbers are small. In PrRu$_4$P$_{12}$, both the increase in $S$
across $T_{\rm MI}$ and the main contribution to the large negative
value of $S$ at low $T$ may be ascribable to this reason, indicating
the change in the Fermi surfaces below $T_{\rm MI}$. The large
negative peak in $S(T)$, suggesting a presence of strong
correlation, is often ascribed to the growth of magnetic correlation
and/or Kondo effect. Such a large $S$ might be of technological
importance as potential thermoelectric material at this low
temperature range. Figure~\ref{Thermo_T}(b) shows the thermoelectric
power factor $S^2/\rho$, which defines the electrical performance of
the thermoelectric material, in PrRu$_4$P$_{12}$. The value of
$S^2/\rho$ increases across $T_{\rm MI}$ tracing the M-I transition,
while it shows a peak around 50 K followed by a steep increase below
$\sim$ 30 K leading to a peak around this temperature and finally it
decreases to zero. The maximum value of $S^2/\rho$
$\sim$2.8~mW/mK$^{-2}$ around 30 K is promising as a thermoelectric
material. One can compare this value at 30 K with that of
Bi$_{86}$Sb$_{14}$ alloy,\cite{Kitagawa} which is known for a good
thermoelectric material.

\subsection{Hall effect}
\begin{figure}[hbtp]
\begin{center}\leavevmode
\includegraphics[width=8.6cm]{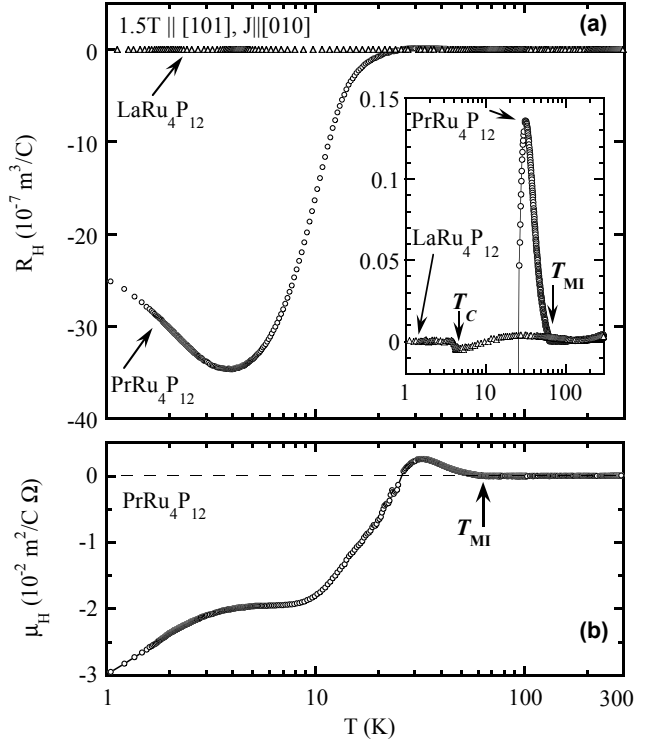}
\caption{(a). Temperature dependence of Hall coefficient $R_{\rm H}$
in PrRu$_4$P$_{12}$ and LaRu$_4$P$_{12}$ for $J\parallel[010]$ and
$H$ = 1.5 T $\parallel [101]$. The inset shows an expanded vertical
axis of the plot. (b) The temperature dependence of Hall mobility
$\mu_{\rm H}$ for the same geometry in PrRu$_4$P$_{12}$.}
\label{Hall_T}\end{center}\end{figure}

Figure~\ref{Hall_T}(a) shows the temperature dependence of the Hall
coefficient $R_{\rm H} (T)$, where the room temperature (RT) values
of $R_{\rm H}\simeq 4 \times 10^{-10}$ and 3.2 $\times
10^{-10}$m$^3$/C yield a carrier concentration of $\sim 1.6 \times
10^{28}$ and $\sim 1.9 \times 10^{28}$/m$^3$ for PrRu$_4$P$_{12}$
and LaRu$_4$P$_{12}$, respectively. These similar values of $R_{\rm
H}$ at RT indicate the similar electronic states in both compounds,
as also indicated by the $S(T)$ behavior. In LaRu$_4$P$_{12}$,
$R_{\rm H}$ shows a weak $T$ dependence with a minimum around 155 K
and a broad peak around $\sim$ 30 K followed by a sign change below
11 K (see the inset), this complex temperature dependence might be
due to the temperature dependence of anisotropy in the relaxation
time.\cite{chambers} Below $\sim$5 K, $R_{\rm H}$ increases and then
goes sharply to zero due to superconductivity below $T_c \simeq$ 4 K
(at 1.5 T) as indicated by the arrow in the inset. In
PrRu$_4$P$_{12}$, $R_{\rm H}$ shows an increase across $T_{\rm MI}$
(clearly visible in the inset) tracing the M-I transition, which
indicate a change in the electronic state. The increase in $R_{\rm
H}$ with decreasing $T$ leads to a maximum around $\sim$30 K below
which $R_{\rm H}$ drops rapidly to change sign to negative and then
shows a minimum around 4 K. Note that $\rho(T)$ also increases below
30 K and shows a peak around 4 K at 1.5 T [see Fig.~\ref{rho_T}(a)].
Below 30 K one of the Pr sties obtains $\Gamma_4^{(2)}$ ground
state, while the other retains $\Gamma_1$ ground state.\cite{iwasa}
Such a minimum in $R_{\rm H}(T)$ is sometimes ascribed to the growth
of antiferromagnetic correlations and/or Kondo effect.
Figure~\ref{Hall_T}(b) shows the Hall mobility $\mu_{\rm H} = R_{\rm
H}/\rho$ at 1.5 T. In a simple one-band picture, $\mu_{\rm H}$ is
directly proportional to the quasi-particle life time and
independent of carrier density. Across $T_{\rm MI}$,  $\mu_{\rm H}$
increases tracing the M-I transition and then rapidly drops below 30
K. It changes sign from positive to negative below 26 K and then
saturates below 10 K followed by a further drop around 4 K. The Hall
conductivity behavior indicates an existence of
temperature-dependent carriers scattering from Pr-sites in the
non-metallic ordered state.

\begin{figure}[hbtp]
\begin{center}\leavevmode
\includegraphics[width=8.6cm]{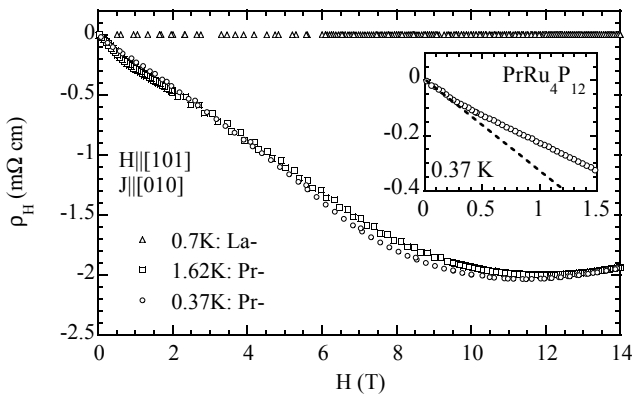}
\caption{. Field dependence of isothermal Hall resistivity
($\rho_{\rm H}$) in (Pr-, La-)Ru$_4$P$_{12}$. The inset shows the
expanded view of $\rho_{\rm H}$ at the low $H$ region in
PrRu$_4$P$_{12}$, where the dotted line indicates a change in
slope.} \label{P_H-H}\end{center}\end{figure}

Figure~\ref{P_H-H} shows the field dependence of Hall resistivity
$\rho_{\rm H}(H)$ in PrRu$_4$P$_{12}$ and LaRu$_4$P$_{12}$ at
several constant temperatures. In both compounds, the sign of
$\rho_{\rm H}$ is negative up to 14 T, although the absolute value
is several orders higher in PrRu$_4$P$_{12}$. There is a clear
change in slope in $\rho_{\rm H}(H)$ at $H\sim0.2$ T in
PrRu$_4$P$_{12}$ (see the inset), where $\rho$ drops to 70\% as
shown in Fig.~\ref{MR-H}(a) and $M$ also increases
sharply.\cite{sakakibara} Tentatively considering the concept of
anomalous Hall effect, i.e., $\rho_{\rm H} (H)=R_0H+R_SM$, our
analysis shows that the anomalous part $R_SM$ can neither be
explained by the skew scattering ($\propto \rho M$) nor by the side
jump scattering ($\propto \rho^2M$) contribution (not shown here).
At higher fields the absolute value of $\rho_{\rm H}$ increases
almost linearly with $H$ up to 6 T and shows a broad peak around 11
T. Taking into account the absence of anomaly in $M(H)$ around this
field range,\cite{sakakibara} this peak indicates an interesting
field-induced change in electronic state.

\section{DICUSSION}
According to the results presented in this paper, the anomalies
across $T_{\rm MI}$ in the electric and thermoelectric transport
indicate that most of the Fermi surface vanishes below $T_{\rm MI}$,
leading to an increase in $\rho$. The observation of broad and sharp
peaks above and below $T_{\rm MI}$, respectively, in the inelastic
neutron scattering (INS) measurement suggests that the 4$f$
electrons shift from itinerant above $T_{\rm MI}$ to rather
localized state below $T_{\rm MI}$.\cite{iwasa} Considering these
facts, the phase transition at $T_{\rm MI}$ could be a hexadecapole
(or higher multipole and scalar) order as argued in the theoretical
model,\cite{takimoto} where no breaking of the local symmetry and
existence of strong $c$-$f$ hybridization have been predicted. The
CDW of the conduction electron associated with this order is an
unconventional one,\cite{shiina} which is triggered by an
interaction with $f$-electrons, unlike the conventional CDW
stabilized by the cooperation between the Fermi-surface instability
and the electron-phonon interaction. The anisotropy in the single
crystal has been found very small in the ordered phase, which is
compatible with the hexadecapole model. However, the resistivity
shows shoulder at intermediate temperatures below $T_{\rm MI}$. In
order to understand this with the other anomalous behaviors observed
in the ordered phase the INS results, suggesting a change in the
strength of hybridization and the CEF level crossing as described in
Sec. I, should be taken into account. The $c$-$f$ hybridization
between the $a_u$ conduction band, which is contributed mainly by
pnictogen $p$ orbits, and the 4$f$ states,\cite{Harima4} combined
with the formation of charge density modulation with the wave vector
$q=$ (1,0,0) due to Fermi-surface nesting, has been ascribed to play
the major role in the modification of the CEF level schemes, i.e.,
CEF level crossing in the low-temperature region.\cite{iwasa1} The
positive-valence Ru-ion displacement closer to Pr$_1$ and further
from Pr$_2$, resulting in the larger and smaller point-charge
Coulomb potentials for the Pr 4$f$ electrons, respectively, also
contributes in the CEF splitting, however this contribution is only
a little because of its small magnitude of the order of $10^{-3}$.
The $c$-$f$ hybridization effect is treated as the perturbation
involving two channels of intermediate states of 4$f^1$ with the
creation of an electron in the vacant states and 4$f^3$ with a hole
in the filled states.\cite{iwasa1} It is demonstrated that the
4$f^3$ process pulls down the $\Gamma_4^{(2)}$ triplet, and the
4$f^1$ process works in an opposite way. Above $T_{\rm MI}$, the
contributions of both the 4$f^1$ and 4$f^3$ processes are
comparable, so that the total effect on the CEF splitting may not be
sufficient to pull down $\Gamma_4^{(2)}$ to a much lower energy
level. With decreasing temperature below $T_{\rm MI}$, the Fermi
surface starts to vanish and the 4$f$ electrons shift to a rather
localized state leading to the different CEF schemes between Pr$_1$
and Pr$_2$.\cite{iwasa1} The contribution of the 4$f^3$ process
becomes larger than that of the 4$f^1$ process at Pr$_2$ and the
contributions of both processes remain comparable at Pr$_1$.

This CEF level crossing could be responsible for the observed
anomalies around 30 K in all measured properties. The level
occupancy of the CEF singlet and triplet is expected to be
fluctuating at least around and above the level crossing
temperature. These (orbital) fluctuations give rise to scattering of
conduction electrons and smears the gap edge of the
conduction-electron density of states (DOS). The suppression of the
gap increases the carrier density reducing the values of $\rho$ in a
region of intermediate temperatures below $T_{\rm MI}$. Below the
level-crossing temperature, $\rho$ shows an upturn again, although
it remains finite even at the lowest temperature, indicating a
low-carrier semi-metallic ground state. A large residual resistivity
$\rho_0$ at $T\sim$0 usually originates from randomness, since
theoretically periodic lattice without randomness should give
$\rho_0\sim$ 0 or infinite value. Randomness can be arisen from
impurity/vacancy or random splitting of the triplet in one of the
sublattices. The present large $\rho_0$ is likely to be originated
by the latter mechanism, since $\rho$ is very sensitive to $H$ only
below 10 K; in particular, $\rho_0$ is drastically decreased by a
small field ($\leq$1 T). Therefore, the scattering of conduction
electrons from random split triplets is suppressed due to the
uniform alignment of the triplet. The above arguments implicitly
assumes an small exchange interaction of the form $H_{ex} =
J_{ex}S_cS_f$, where $J_{ex}$ is the exchange constant corresponding
to the so-called on-site Kondo coupling, $S_c$ is the
conduction-electron spin, and $S_f$ is the 4$f$ electron spin
populated at the triplet $\Gamma_4^{(2)}$. Formally, if $S_f$ is
random, conduction electron has to move in a random (magnetic)
potential. On the other hand, the potential becomes uniform when all
$S_f$ align along one direction. The value of $J_{ex}$, accounting
the hybridization between $p$-state of the cage ions and $f$-state
at the center, is expected to be of the order of 1 K, since only the
low-temperature ($\leq$ 10K) resistivity is influenced by the
applied magnetic fields. The absence of any magnetic order even down
to 20 mK has been confirmed by the $\mu$SR measurement.\cite{saha4}
Thus, the obtained effect of magnetic fields, particularly GMR at
low $T$, observed only in Pr compound, could be arisen due to the
suppression of the Kondo scattering of the charge carriers from the
4$f$-electrons. Finally, under high magnetic fields the transverse
magnetoresistance is anisotropic and the Hall resistivity shows a
broad peak, despite no anomaly and little anisotropy in the
magnetization at high fields. These facts indicate a magnetic field
induced change in electronic state, the detail of which would be
interesting for further investigation.

\section{SUMMARY}
In summary, we have investigated  the transport and magnetic
properties in single-crystalline PrRu$_4$P$_{12}$. The phase
transition below 63 K might be due to an antiferro-hexadecapole
order where a dominant $c$-$f$ hybridization plays an essential
role. An additional structure has been recognized below 30 K
exhibiting the anomalous behaviors, e.g., highly temperature
dependent carrier scattering, giant negative magnetoresistance,
large themopower $\sim -200\mu$V/K, etc., without showing any
magnetic order. Judging from the resistivity behavior, the ground
state emerges as a low-carrier semimetal. These anomalous behaviors
are most likely associated with the $c$-$f$ hybridization between
the remaining charge carriers and Pr 4$f$ electrons, resulting in a
change in CEF ground state of the Pr$_2$ sites at the cubic corner
from $\Gamma_1$ to $\Gamma_4^{(2)}$.

\vskip 0.5cm
\begin{center}
{ACKNOWLEDGEMENTS}
\end{center}
We thank R. Shiina, H. Harima, K.~Iwasa, and C.~Sekine for useful
discussions. One of the authors (S.R.S) acknowledges the support
from the Japan Society for the Promotion of Science (JSPS) under
Grant No. P01028. This work was partly supported by a Grant-in-Aid
for Scientific Research Priority Area "Skutterudite" (Grant
No.15072206) of the Ministry of Education, Culture, Sports, Science
and Technology of Japan.


\end{document}